\documentclass[lettersize,journal]{IEEEtran}
\usepackage{amsmath,amsfonts}
\usepackage{algorithmic}
\usepackage{algorithm}
\usepackage{array}
\usepackage[caption=false,font=normalsize,labelfont=sf,textfont=sf]{subfig}
\usepackage{textcomp}
\usepackage{stfloats}
\usepackage{url}
\usepackage{verbatim}
\usepackage{graphicx}
\usepackage{cite}
\begin{document}

\title{Integrated Channel Sounding and Communication: Requirements, Architecture, Challenges, and Key Technologies}

\author{Nanhao Zhou, Chao Zou, Yu Zhou, Yanqun Tang, Xiaoying Zhang, Haoran Yin, Xuefeng Yin, \textit{Member, IEEE}, Yuxiang Zhang, \textit{Member, IEEE}, Dan Fei, and Fan Jiang, \textit{Member, IEEE}
\thanks{This work was supported by the Shenzhen Science and Technology Major Project under Grant KJZD20240903102000001, in part by the Science and Technology Planning Project of Key Laboratory of Advanced IntelliSense Technology, Guangdong Science and Technology Department under Grant 2023B1212060024. \textit{(Corresponding authors: Yanqun Tang; Xiaoying Zhang.)}}
\thanks{Nanhao Zhou, Chao Zou, Yu Zhou, and Haoran Yin are with the School of Electronics and Communication Engineering, Sun Yat-sen University, China (email: zhounh3@mail2, zouch28@mail2, zhouy633@mail2,  tangyq8@mail, yinhr6@mail2).sysu.edu.cn}
\thanks{Yanqun Tang is with the School of Electronics and Communication Engineering, Sun Yat-sen University, China, and also with the Guangdong Provincial Key Laboratory of Sea-Air-Space Communication, China (email:  tangyq8@mail.sysu.edu.cn}
\thanks{Xiaoying Zhang is with the College of Electronic Science and Technology,
National University of Defense Technology, China (email: zhangxiaoying@nudt.edu.cn).}
\thanks{Xuefeng Yin is with the College of Electronics and Information Engineering, Tongji University, China (email: yinxuefeng@tongji.edu.cn).}
\thanks{Yuxiang Zhang is with the State Key Laboratory of Networking and Switching Technology,
Beijing University of Posts and Telecommunications, China (email: zhangyx@bupt.edu.cn).}
\thanks{Dan Fei is with the School of Electronic and Information Engineering, Beijing Jiaotong University, China (email: dfei@bjtu.edu.cn).}
\thanks{Fan Jiang is with the Department of Broadband Communication, Pengcheng Laboratory, China (email:  jiangf02@pcl.ac.cn).}
%\thanks{Fan Liu is with the National Mobile Communications Research Laboratory, Southeast University, China (email:  f.liu@ieee.org).}
}

% The paper headers
% \markboth{Journal of \LaTeX\ Class Files,~Vol.~14, No.~8, August~2021}%
% {Shell \MakeLowercase{\textit{et al.}}: A Sample Article Using IEEEtran.cls for IEEE Journals}

% \IEEEpubid{0000--0000/00\$00.00~\copyright~2021 IEEE}
% Remember, if you use this you must call \IEEEpubidadjcol in the second
% column for its text to clear the IEEEpubid mark.

\maketitle

\begin{abstract}
Channel models are  essential for the design, evaluation, and optimization of wireless communication systems. The emerging space–air–ground–sea integrated network (SAGSIN), characterized by diverse service applications and extended-spectrum operations, places even greater demands on highly accurate channel models. However, conventional channel sounding is limited by generalized measurement campaigns, inadequate cross-band consistency, and insufficient real-time adaptability, making it unable to meet the needs of SAGSIN for scenario-specific and high-precision channel modeling. To address this challenge, we propose a novel technological framework, termed integrated channel sounding and communication (ICSC). By deeply integrating sounding and communication, the ICSC enables efficient and real-time acquisition of dynamic channel characteristics during communication processes, supporting fine-grained site- and scenario-specific measurements. Furthermore, leveraging artificial intelligence techniques, ICSC can identify channel conditions and adapt waveform parameters in real-time according to scenario variations, which in turn enhances communication performance. This article first introduces the fundamental principles of the ICSC framework, elaborates on its core concepts and key advantages, and demonstrates its feasibility through the development of an integrated verification system (IVS). Subsequently, the potential applications and opportunities of the ICSC are analyzed in depth, followed by a discussion of its future development directions and remaining challenges.

\end{abstract}

% \begin{IEEEkeywords}
%  ICSC, channel sounding and modeling, scenario identification, communication enhancement.
% \end{IEEEkeywords}

\section{Introduction}
Channel models characterize wireless signal propagation in the physical environment by quantifying key parameters, such as shadow fading (SF), path loss (PL), delay spread (DS), angular spread (AS), and Doppler shift. As the medium for electromagnetic wave propagation, the wireless channel is collectively determined by the geometry, material properties, and dynamic evolution of environmental scatterers. Consequently, the primary objective of channel modeling is to identify and quantify the dominant propagation mechanisms in complex dynamic environments, as well as to parameterize channel characteristics. Current mainstream channel modeling approaches can be roughly categorized into empirical, deterministic, and semi-deterministic methods\cite{ref1}. Empirical models, typically derived from large-scale measurement datasets, have low computational complexity and offer simple, and intuitive formulations. However, due to the coarse partitioning of scenarios, their predictions may deviate substantially from the actual channel characteristics in certain environments. Deterministic modeling, on the contrary, is based on precise environmental information and electromagnetic calculations\cite{ref2}. Nevertheless, despite its high accuracy, deterministic modeling faces persistent challenges due to high computational complexity. Semi-deterministic modeling offers a balanced compromise by appropriately abstracting and simplifying environmental information, thereby achieving a practical trade-off between complexity and accuracy\cite{ref3}. 

%, and a complete recomputation is required for each unmodeled scenario

%\begin{figure}[!t]
%\centering
%\includegraphics[width=0.5\textwidth]{Fig1.pdf}
%\caption{An illustration of space-air-ground-sea integrated networks.}
%\label{fig3}
%\end{figure}
\par In real-world applications, the complexity of the propagation environment makes it difficult for mainstream modeling methods to balance simplification and accuracy effectively. Excessive simplification undermines the ability of the model to accurately represent propagation mechanisms, while incorporating all scatterers significantly increases computational complexity\cite{ref4}. Additionally, the future space-air-ground-sea integrated network (SAGSIN), which aims to support the deep integration of diverse services and extended-spectrum operations, presents new and challenging demands for channel modeling. In SAGSIN, emerging communication scenarios display high dynamics (e.g., high-speed rail, unmanned aerial vehicles, and intelligent connected vehicles), complex geographical environments (e.g., mountainous terrain, canyons, and tunnels), and unique propagation characteristics in specialized environments (e.g., underwater communication). Meanwhile, to meet diverse service demands, SAGSIN must support multi-band cooperation, including Sub-6 GHz, millimeter-wave (mmWave), and terahertz (THz) bands\cite{ref5}. Notably, channel propagation characteristics vary markedly across different frequency bands. Sub-6 GHz channels exhibit pronounced fading, abundant reflections, and relatively low path loss. The mmWave channels suffer from high penetration loss and sparse scattering, while THz channels are dominated by strong molecular absorption and weak multipath effects\cite{ref6,ref7,ref8}. Additionally, the diversity of wireless technologies further exacerbates the complexity of modeling requirements. For example, massive multiple-input multiple-output (MIMO) requires investigation of angular spread and spatial correlation, while beamforming requires analysis of angular distribution, array gain, and directional losses\cite{ref9}. Consequently, channel modeling for SAGSIN requires extensive sounding platforms that span various scenarios, frequency bands, and technologies. Such platforms, however, rely on specialized equipment, incur substantial costs, offer limited coverage, and are poorly adaptable to multi-band cooperative scenarios. As a result, the collected data may be insufficient to support the precise development of SAGSIN channel models, further restricting the scalability and applicability of traditional modeling methods.

\par Therefore, new sounding techniques are urgently needed to support the complex and heterogeneous scenarios anticipated in SAGSIN. Motivated by this necessity, we propose integrated channel sounding and communication (ICSC), a unified framework that combines wireless communication and sounding functionalities. By opportunistically using communication signals for sounding, ICSC overcomes the limitations of traditional sounding methods that depend on dedicated probing sequences and specialized sounding platforms. It enables the acquisition of channel state information (CSI) during data transmission, achieving tight integration between communication and sounding. Leveraging existing communication infrastructure, ICSC supports large-scale, continuous channel sounding across multiple domains, including space, air, ground, and sea, at substantially lower cost than dedicated platforms, effectively facilitating modeling across diverse scenarios and frequency bands. By virtue of being communication-driven, ICSC inherently provides high temporal resolution and broad spatial coverage, enabling the system to capture instantaneous, high-precision channel characteristics in diverse environments. These capabilities provide a rich and reliable data foundation for channel modeling and prediction in complex scenarios. Moreover, by harnessing existing communication infrastructure, ICSC enables finer-grained site- and scenario-specific classification sounding and delivers more detailed channel sounding than traditional methods. This capability not only mitigates the inaccuracies caused by coarse scenario partitioning in conventional channel modeling, where model outputs may deviate significantly from actual channel conditions, but also retains the low computational complexity of empirical models.

\begin{figure*}[!t]
\centering
\includegraphics[width=0.9\textwidth, height=0.8\textheight]{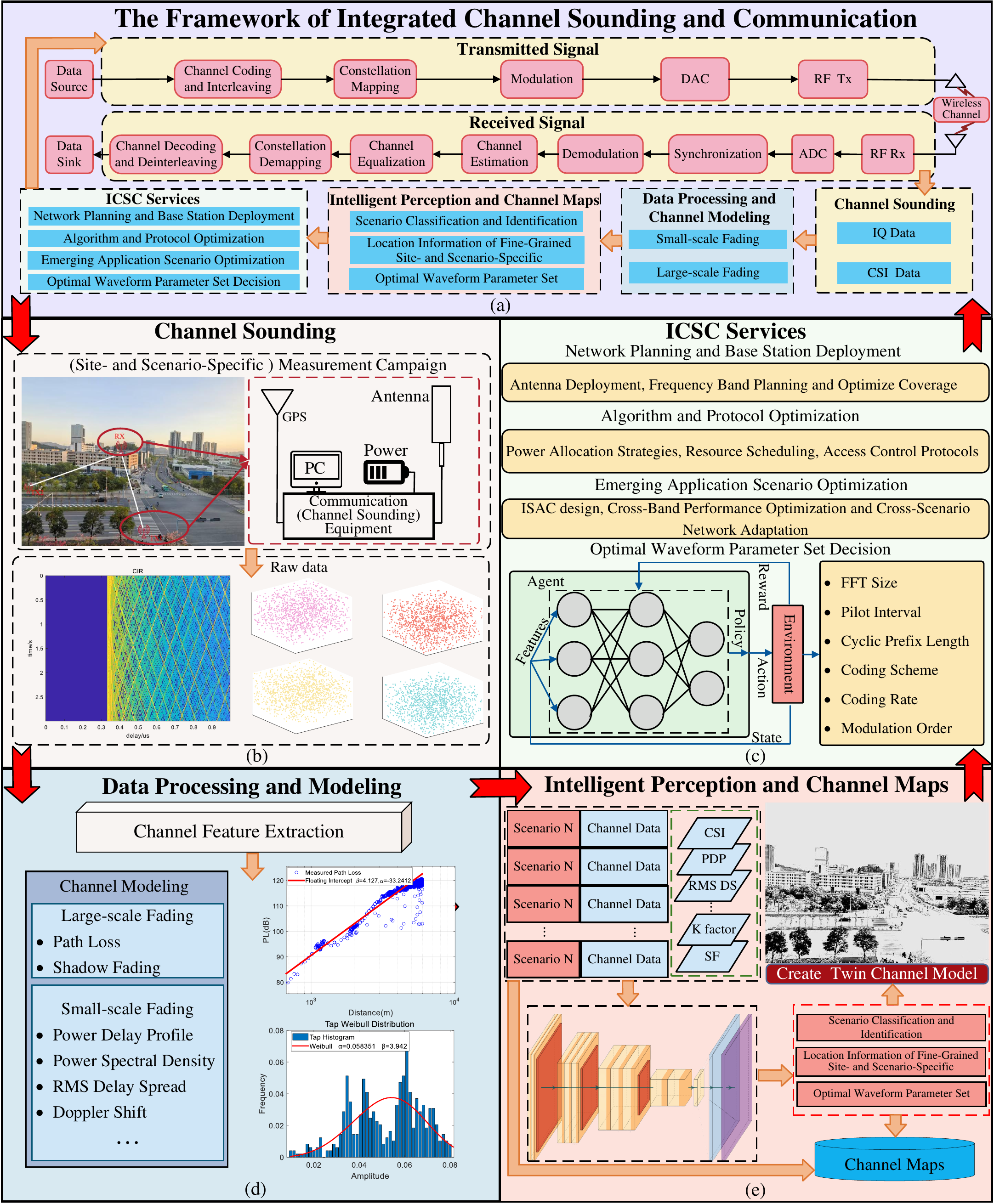}
\caption{The framework of integrated channel sounding and communication.}
\label{fig2}
\end{figure*}

\par The rest of this article is organized as follows. We first detail the fundamental principles of the ICSC framework and summarize its main advantages. Then an integrated verification system (IVS) is developed to demonstrate the feasibility of the proposed framework. In the end, potential applications, future research directions, and technical challenges are discussed.

\section{Integrated Channel Sounding and Communication Framework}
ICSC enables the simultaneous transmission of information and channel sounding via shared resources, which not only improves spectrum efficiency and overall system performance but also enhances adaptability to various operational scenarios. This section introduces the fundamental principles of ICSC and highlights its key advantages.

\subsection{Overall Principles}
The core principle of ICSC is to use communication signals for the simultaneous transmission of information and channel sounding, thus achieving a deep integration of communication and measurement functionalities. Unlike traditional methods that rely on dedicated sounding sequences and independent measurement systems, ICSC directly extracts, statistically analyzes, and exploits inherent features of service signals, such as multipath delay, Doppler shift, and angle information, during transmission to conduct channel sounding. This enables real-time, continuous channel sounding across diverse scenarios within a SAGSIN, free from additional overhead. More importantly, ICSC facilitates the efficient establishment of comprehensive channel model libraries that encompass fine-grained site- and scenario-specific conditions across multiple frequency bands. By integrating intelligent environment perception and deep learning (DL)-driven dynamic waveform optimization, ICSC maintains optimal communication strategies while balancing performance gains and system costs. Additionally, ICSC generates multi-dimensional channel data that provides reliable support for network planning, protocol design, and the deployment of emerging technologies. The ICSC framework is illustrated in Fig. 1, and the specific procedure is followed. 

\subsubsection{Communication}
As shown in Fig. 1(a), at the transmitter, original data bits are encoded and mapped onto a modulation constellation to generate the baseband signal. This signal is subsequently modulated and radiated by the antenna. During propagation, the signal is subject to channel fading and interference. At the receiver, the received signal undergoes synchronization and demodulation, with CSI acquired through channel estimation. Using the estimated CSI, the receiver performs channel equalization followed by decoding to recover the original data bitstream.

\subsubsection{Channel Sounding and Modeling}
As illustrated in Fig. 1(b) and Fig. 1(d), during communication, the receiver actively acquires real-time CSI through data demodulation, enabling active channel sounding. Specifically, during data demodulation, in-phase and quadrature (IQ) samples and CSI are collected and analyzed via high-resolution channel parameter extraction algorithms, such as the space-alternating generalized expectation-maximization (SAGE). This analysis yields both large-scale fading characteristics (e.g., PL and SF) and small-scale parameters (e.g., power delay profile (PDP), angles of arrival (AOA), and root mean square DS (RMS DS)). Based on these measurements, fine-grained site- and scenario-specific channel models can be developed to accurately characterize the propagation environment, providing reliable support for intelligent scenario identification (SI) and adaptive waveform optimization. Notably, traditional channel sounding methods typically obtain only downlink characteristics directly, while uplink properties are inferred from channel reciprocity. However, such inferred parameters may differ substantially from the actual propagation conditions. In contrast, the ICSC approach allows the simultaneous acquisition of both uplink and downlink channel characteristics during communication, thereby overcoming the limitations of conventional sounding techniques and significantly enhancing the completeness and accuracy of channel modeling.

\subsubsection{Intelligent Perception and Channel Maps}
ICSC takes advantage of ubiquitous communication networks to continuously obtain abundant channel data. Furthermore, uploading the receiver’s geographic location information together with measured channel characteristics to a base station or cloud database facilitates the development of a large-scale channel model library, enables the generation of channel maps, and establishes a robust data foundation for comprehensive regional channel characterization. Additionally, using real-time CSI, IQ data and other channel features, the system applies artificial intelligence (AI) algorithms (e.g., convolutional neural networks (CNNs) and support vector machines (SVMs)) to perform inference, thus enabling dynamic scenario classification and identification.

\subsubsection{ICSC Services}
In real-time, the system leverages measured CSI together with intelligent perception outputs to enable DL models to rapidly determine optimal waveform parameters, such as modulation scheme, coding rate, cyclic prefix length, and fast Fourier transform (FFT) size, etc., as presented in  Fig. 1(c). The selected parameters are fed back to the transmitter for dynamic adaptation, facilitating waveform adjustment and efficient resource allocation. As a result, this adaptive mechanism helps alleviate performance degradation caused by multipath propagation, interference, and fading, thereby ensuring stable communication while concurrently reducing overall system overhead. Offline, the data and models stored in the model library support network planning and base station deployment. Furthermore, these models can be used to evaluate system performance and optimize algorithms and protocols, thus improving the robustness and efficiency of the system. In addition, multidimensional channel features obtained through the ICSC provide data-driven support for emerging applications, including integrated sensing and communication (ISAC), cross-scenario optimization for new 6G frequency bands, and SAGSIN-related technologies.

\subsection{Advantages of ICSC}
The proposed ICSC method offers four key advantages:

\subsubsection{Acquisition of Real-Time CSI Data}

The ICSC demonstrates robust real-time CSI extraction and prediction capabilities, enabling adaptation to complex and dynamic communication environments. Consequently, it maintains stable operation across scenario variations and provides a reliable stream of high-quality channel data.

\subsubsection{Fine-Grained Site- and Scenario-Specific Channel Sounding}
ICSC utilizes widely deployed communication networks to perform large-scale, continuous channel sounding across multiple domains (space, air, ground, and sea) in real-time and at low cost. This strategy enables the comprehensive acquisition of channel characteristics, supports finer-grained, scenario-specific sounding, and thus effectively overcomes the limitations of conventional sounding methods.

\subsubsection{Real-Time Scenario Identification and Classification}
Building on CSI and other real-time data acquired via ICSC, data-driven AI methods can rapidly identify and classify communication scenarios, establish precise mappings between the environment and channel characteristics, and thereby facilitate the generation of accurate channel maps.

\subsubsection{Real-Time Optimization of Communication Performance}
ICSC determines optimal communication strategies in real-time based on SI outcomes and measured CSI, thereby ensuring robust and reliable communication during transitions across complex, time-varying scenarios.

\section{Demonstration of the Integrated Validation System for ICSC}
To verify the feasibility of the ICSC architecture, an IVS is developed specifically for the technology. The system integrates several key functions, including communication performance evaluation, channel sounding, intelligent SI, and adaptive decision-making driven by waveform parameters. When a scenario change is identified by the system through CSI acquired via real-time channel sounding, the system uses DL algorithms to determine the optimal waveform parameters, after which the transmitter adjusts the waveform parameters accordingly to maintain stable communication performance in dynamic environments.

\begin{figure}[H]
\centering
\includegraphics[width=0.48\textwidth]{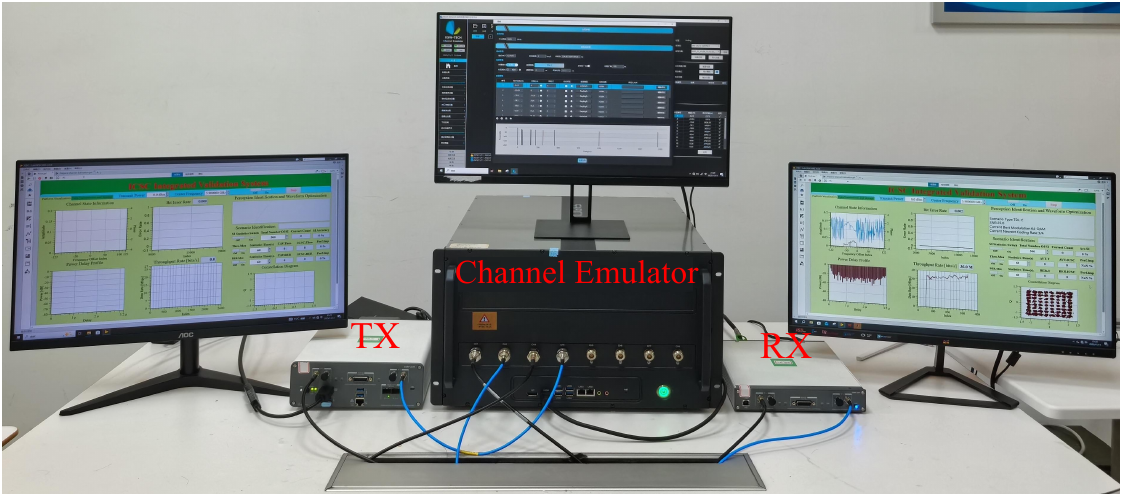}
\caption{The integrated validation system for ICSC architecture.}
\label{fig2}
\end{figure}

\par This section presents the design and implementation of the IVS tailored to the ICSC, as illustrated in Fig. 2. The hardware platform comprises two universal software radio peripheral devices (USRP), while the software environment is developed in LabVIEW NXG with an integrated visualization interface. The system establishes a complete end-to-end communication link, enabling real-time acquisition of CSI and statistical analysis of channel characteristics. The experiment utilized a system design based on the IEEE 802.11ac standard, with a carrier frequency of 5.9 GHz, a bandwidth of 20 MHz, and 64 subcarriers. The set of waveform-parameters included nine modulation and coding schemes (MCS): BPSK(1/2), QPSK(1/2), QPSK(3/4), 16-QAM(1/2), 16-QAM(3/4), 64-QAM(2/3), 64-QAM(3/4), 64-QAM(5/6), and 256-QAM(3/4). In terms of intelligent processing, the system integrates an SI module based on CNNs and a waveform parameter decision module guided by reinforcement learning (RL). Due to their local perception and parameter sharing properties, CNNs demonstrate strong generalization and robustness for complex pattern recognition tasks. The CNNs in this model comprise 5 convolutional layers, each with 32 filters and a kernel size of 9\cite{ref10}. The batch size is set to 100, 50 epochs are trained, and the dimensionality of the channel is 53 × 2. The decision module adopts the dueling double deep Q-network (D3QN) to intelligently select the optimal waveform parameters based on SI results and current communication quality (e.g., signal-to-noise ratio (SNR)). The D3QN architecture incorporates a 64 dimensional feature extraction layer and separate value and advantage branches, each containing 64 dimensional fully connected layers. In D3QN network, the agent observes the communication scenario type and SNR as the state, selects a MCS as the action, and receives a reward according to whether the selected configuration achieves the maximum achievable throughput under the prevailing channel conditions. The batch size is set to 128, 1500 episodes are trained. Following scenario change, the system rapidly determines an optimal set of waveform parameters, completes the waveform switching, and ensures the stability of the communication link. 
%The experiments utilized an orthogonal frequency division multiplexing (OFDM) waveform with a carrier frequency of 5.9 GHz, a bandwidth of 20 MHz, and 64 subcarriers.

\begin{figure*}[t]
\centering
\includegraphics[width=1\textwidth]{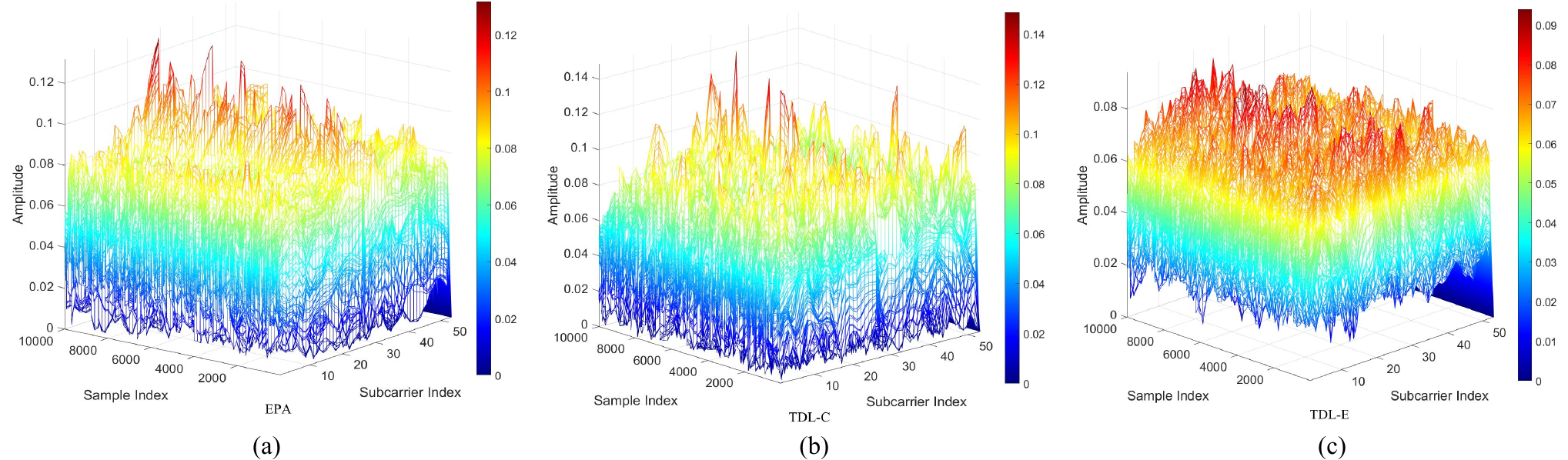}
\caption{The visualization of different channels.  (a)EPA channel. (b)TDL-C channel. (c)TDL-E channel.}
\label{Fig4}
\end{figure*}

 \begin{figure}[H]
\centering
\includegraphics[width=0.45\textwidth, height=0.45\textheight]{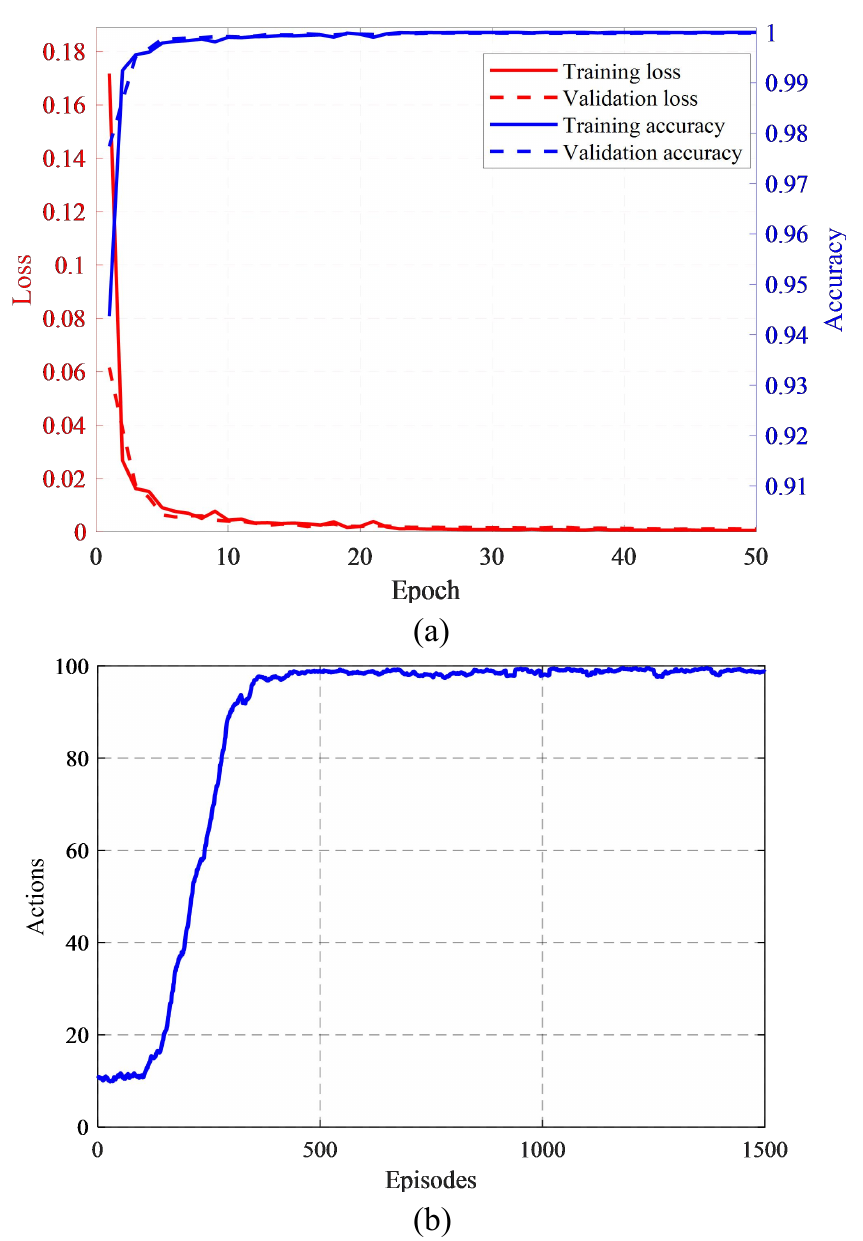}
\caption{Training results. (a)The SI training results. (b)The waveform decision training results.}
\label{fig3}
\end{figure}

\par To enable controlled evaluation across various communication scenarios, a channel emulator was used to emulate the physical wireless channel, and three representative 3GPP standard channel models, EPA, TDL-C, and TDL-E, were adopted to emulate different communication scenarios, facilitating real-time switching during transmission.

\par The experimental results of the ICSC IVS are presented in Figs. 3, 4, and 5, with the corresponding statistical results summarized in Table 1. Fig. 3(a)–(c) illustrate the visualization of sounding results for different channels. Fig. 4(a) shows the training performance of the SI, while Fig. 4(b) presents the waveform decision training results. Fig. 5 shows the overall verification results of ICSC for the TDL-E channel. The IVS provides real-time visualization of communication and measurement data—including CSI, PDP, data rates, and constellation diagrams. Upon channel switching, ICSC accurately identifies the channel type, rapidly determines the optimal waveform parameters, and executes timely adjustments. Descriptions of the table entries  as follows: Total Number of SI (TN.SI), Correct Count (Corr.C), average throughput of the traditional algorithm (AVT.T), average throughput of ICSC (AVT.ICSC), and performance improvement in throughput (Perf.Imp). The unit of throughput is Mbps.

\begin{figure}[H]
\centering
\includegraphics[width=0.48\textwidth]{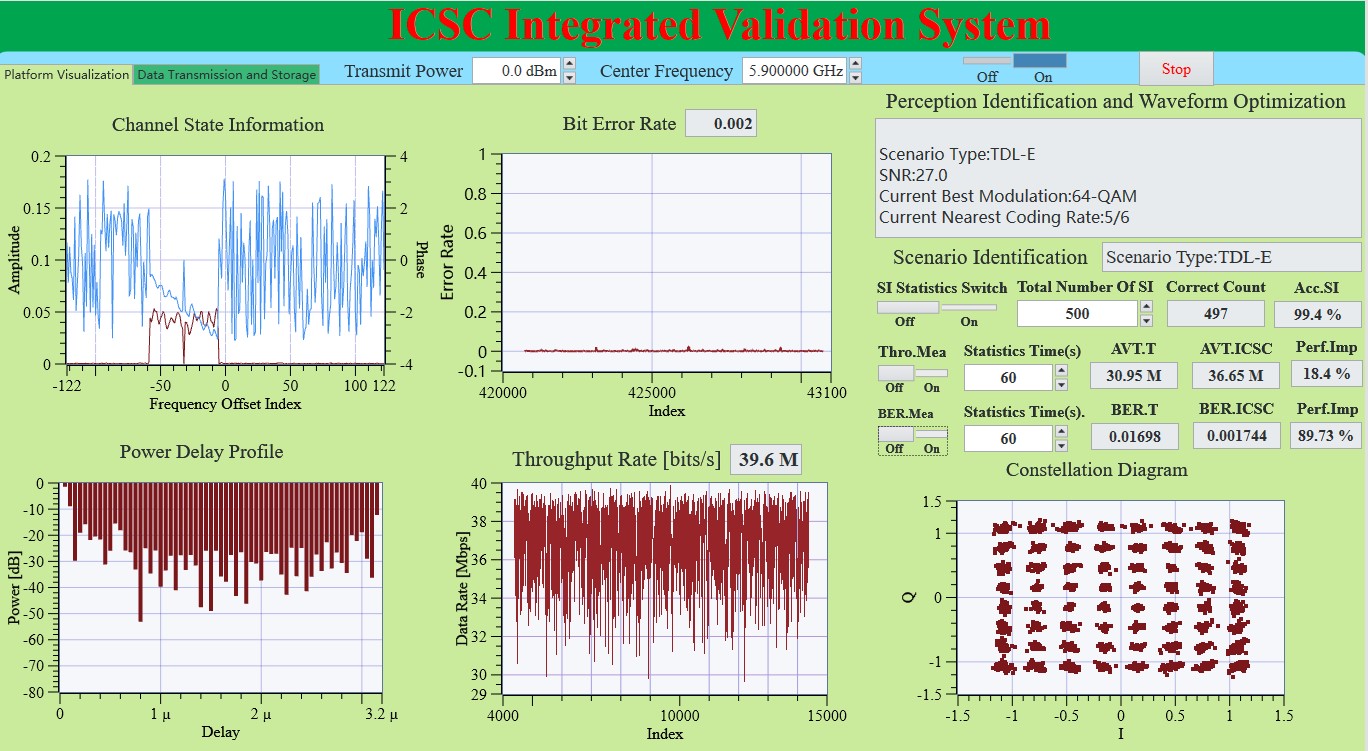}
\caption{The overall verification results of ICSC for the TDL-E channel.}
\label{fig5}
\end{figure}

\par The results indicate that the IVS accurately measures and captures CSI and reliably identifies the corresponding communication scenarios across three representative channel models. In a statistical experiment of 500 SI trials, the identification accuracy for each scenario exceeded 98\%. The ICSC-based waveform decision method efficiently selects the optimal waveform parameters and rapidly adapts to them as communication scenarios change. With respect to average throughput, the ICSC-based approach achieves at least a 10\%\ performance improvement over conventional SNR-based adaptive coding algorithms, while not increasing the average bit error rate (BER). Collectively, these results convincingly demonstrate the feasibility and effectiveness of ICSC technology.

\renewcommand{\arraystretch}{1.5} % 调整行高
\setlength{\tabcolsep}{8pt}     % 调整列间距

\begin{table}[H]
\centering
\caption{ICSC experiment statistical results}

% ======================= 表格内容 =======================
\begin{tabular}{
 >{\centering\arraybackslash}p{0.9cm}
 >{\centering\arraybackslash}p{0.8cm}
 >{\centering\arraybackslash}p{0.8cm}
 >{\centering\arraybackslash}p{0.8cm}
 >{\centering\arraybackslash}p{1.1cm}
 >{\centering\arraybackslash}p{0.9cm}
}

\hline
\textbf{Scenario} & \textbf{TN.SI} & \textbf{Corr.C} & \textbf{AVT.T} & \textbf{AVT.ICSC} & \textbf{Perf.Imp} \\ 
\hline
EPA   & 500 & 494 & 15.27 & 17.56 & 14.9\% \\ 
TDL-C & 500 & 498 & 19.50 & 21.93 & 12.5\% \\ 
TDL-E & 500 & 497 & 30.95 & 36.65 & 18.4\% \\ 
\hline

\end{tabular}
\end{table}

\section{Potential Applications and Opportunities}
%Using the deep integration of communication and channel sounding, the ICSC framework facilitates continuous channel modeling and intelligent communication optimization across multiple scenarios and frequency bands. It also establishes a robust data foundation for network planning, protocol design, and cross-scenario adaptation. 
This section examines the potential applications and opportunities enabled by the key advantages of the ICSC framework.

\subsection{Channel Knowledge Maps}
Channel knowledge maps (CKMs) are databases indexed primarily by terminal or virtual positions. They integrate historical channel data contributed by terminals in a target region, enabling accurate capture of local channel characteristics and providing rich environmental priors\cite{ref11,ref12}. This capability reduces the need for frequent online environmental perception and channel estimation, substantially reducing the computational and communication resources required to acquire CSI. Existing approaches to building CKMs struggle to accurately characterize complex channel behaviors and to adapt to dynamic environments, making efficient and precise CKM construction an ongoing challenge. The ICSC framework, with its ability to perform real-time high precision CSI acquisition and continuous channel modeling, effectively overcomes the limitations of conventional CKMs in capturing channel characteristics under complex propagation conditions. Moreover, ICSC supports the optimization of real-time communication performance by delivering waveform parameters that determine the optimal transmission strategy at each location, thereby contributing to more efficient CKM construction.

\subsection{AI-Enabled Data-Driven Channel Modeling}
Empirical statistical channel models are constrained by the finite number of scenario types and fixed structures, making it difficult to capture the rapid evolution of the channel in complex environments. With the continuous advancement of DL, data-driven AI technologies have emerged as the core solution to this issue. Through the continuous acquisition of high resolution spatio-temporal–frequency (STF) channel data from ICSC in large areas, AI can learn more realistic and comprehensive environmental information, promoting a deeper analytical insight and prediction of dynamic channel characteristics. Powered by DL’s advantages in nonlinear modeling and multi-layer feature extraction, the ICSC-enabled AI channel modeling is expected to reveal implicit relationships between the environment and channels. This capability is expected to address and potentially transcend the limitations of traditional methods in feature extraction, scenario description, and cross-scenario generalization, achieving higher precision and more robust channel modeling and prediction.

\subsection{Environment Intelligence Communication}
Environmental intelligent communication (EIC) acquires wireless environmental information through sensing and employs AI to predict channel fading based on real-time data\cite{ref13}. Such predictions enable autonomous optimization of air interface transmission, facilitating intelligent coordination and adaptive interaction between the communication system and the environment. The ICSC leverages high resolution channel sounding capabilities across multiple scenarios to provide essential channel data support for EIC. Moreover, by identifying real-time changes in communication scenarios and collecting environmental information, ICSC can predict channel states and adaptively adjust air interface configurations (e.g., modulation, coding, and bandwidth) based on those predictions to optimize communication performance. This approach overcomes the limitations of traditional static models, which respond sluggishly to dynamic environments, and enables fine-grained characterization of multipath propagation and environmental variations in complex, heterogeneous scenarios, thereby significantly improving system reliability and robustness and providing a feasible path for EIC.

\section{Challenges and Future Research Directions}
Although the ICSC provides extensive capabilities in channel sounding and communication performance optimization, some critical challenges remain. This section explores the open research questions associated with the ICSC in future communication networks. 

\subsection{Novel Waveforms Design}
With the continuous advancement of future communication toward higher data rates and more complex application scenarios, traditional OFDM waveforms experience performance degradation in the spatial, temporal, and frequency domains due to channel non-stationarity. In particular, under high-mobility and wideband conditions, OFDM is highly susceptible to Doppler spread, which undermines system stability and reliability. This shortcoming imposes more stringent requirements on channel sounding accuracy and transmission reliability within ICSC systems. To address these issues, the design of next-generation waveforms has emerged as a critical research direction. Among the leading candidates, affine frequency division multiplexing (AFDM), which is based on the discrete affine Fourier transform (DAFT), flexibly maps symbols onto a set of orthogonal chirp waveforms. AFDM demonstrates significant strong resilience to doubly selective channels and robustness against carrier frequency offset, while retaining backward compatibility with OFDM and offering low implementation complexity\cite{ref14,ref15}. Deep integration of AFDM with ICSC enables high-precision channel sounding and adaptive communication coordination, while concurrently enhancing spectral efficiency and intelligent sensing in future communication. Building upon this new waveform, the design of frame structures is crucial for implementing ICSC. Its core involves the principled allocation of resources between communication and measurement symbols, ensuring reliable communication while meeting sounding requirements. To meet diverse system demands and dynamic environments, resource allocation must be highly adaptive and dynamically optimized according to real-time channel conditions. 

\subsection{Channel Estimation Optimization}
The overall performance of a communication system is tightly coupled with the accuracy of CSI, which directly dictates the operational effectiveness of ICSC. Under favorable channel conditions, parameter estimation achieves high precision; in contrast, degraded communication environments can introduce substantial estimation error. This issue is exacerbated in complex and dynamic wireless scenarios, where the received signal is simultaneously distorted by timing offset (TO), carrier-frequency offset (CFO), and multipath fading. These detrimental factors not only jointly aggravate estimation errors, but also exhibit strong mutual coupling: TO disrupts symbol orthogonality, thereby degrading the accuracy of both CFO and channel estimation; CFO further undermines the reliability of channel estimation, while the time-varying characteristics of the channel considerably increase the complexity of synchronization compensation. Conventional iterative methods execute timing synchronization, CFO estimation, and channel estimation in sequence, leading to high computational complexity, limited real-time responsiveness, and cumulative error propagation, while failing to exploit the intrinsic correlations among these parameters. Consequently, the joint estimation and compensation of TO, CFO, and multipath fading under real-time constraints have emerged as a pivotal challenge to break through the performance bottlenecks of ICSC systems.

\subsection{Channel Modeling Based on Multimodal Data}
As communication environments grow increasingly complex, traditional channel modeling methods that rely on a single data source can no longer meet the high precision modeling requirements. By integrating antenna parameters, environmental context, terminal states, and device-related factors, multimodal channel modeling delivers a holistic system perspective and markedly strengthens the accuracy and resilience of channel representation. Antenna-related information may comprise cross-polarization ratios, isolation characteristics, and inter-element spacing; environmental factors may span terrain configuration, architectural structures, and atmospheric conditions; and terminal states may include positional and velocity information. Device-related factors, such as thermal noise, phase noise, and self-interference, serve as key multimodal inputs that mitigate hardware deficiencies and enhance model fidelity. Consequently, channel models derived from multimodal data fusion show improved adaptability in dynamic conditions, ensuring a more reliable realization of ICSC capabilities.

\section{Conclusions}
This article presents the ICSC framework that utilizes communication signals for channel sounding and incorporates intelligent perception to achieve real-time identification and classification of channel characteristics. Building on this framework, channel maps can be generated from environmental scenarios and their associated channel features, while simultaneously providing decision support for communication performance optimization. This study systematically describes the design methodology and key advantages of the ICSC framework, highlighting its ability to acquire channel features in real-time and optimize communication performance. Furthermore, the proposed framework is validated through an IVS demonstration, confirming its technical feasibility. This work is expected to stimulate further research and development on ICSC technologies, enhance their adaptability and performance in dynamic and complex network environments, and lay the groundwork for the evolution of next-generation intelligent communication systems.

% \newpage

% \section{Biography Section}
% If you have an EPS/PDF photo (graphicx package needed), extra braces are
%  needed around the contents of the optional argument to biography to prevent
%  the LaTeX parser from getting confused when it sees the complicated
%  $\backslash${\tt{includegraphics}} command within an optional argument. (You can create
%  your own custom macro containing the $\backslash${\tt{includegraphics}} command to make things
%  simpler here.)
 
% \vspace{11pt}

% \bf{If you include a photo:}\vspace{-33pt}
% \begin{IEEEbiography}[{\includegraphics[width=1in,height=1.25in,clip,keepaspectratio]{fig1}}]{Michael Shell}
% Use $\backslash${\tt{begin\{IEEEbiography\}}} and then for the 1st argument use $\backslash${\tt{includegraphics}} to declare and link the author photo.
% Use the author name as the 3rd argument followed by the biography text.
% \end{IEEEbiography}

% \vspace{11pt}

% \bf{If you will not include a photo:}\vspace{-33pt}
% \begin{IEEEbiographynophoto}{John Doe}
% Use $\backslash${\tt{begin\{IEEEbiographynophoto\}}} and the author name as the argument followed by the biography text.
% \end{IEEEbiographynophoto}

\vfill

\end{document}